\documentclass[12pt,preprint]{aastex}

\slugcomment{Version: \today}

\shorttitle{V1647~Orionis in 2011}
\shortauthors{Aspin}

\usepackage{natbib}
\begin{document}

\title{THE CONTINUING OUTBURST OF V1647~ORIONIS: WINTER/SPRING 2011
  OBSERVATIONS}

\author{
Colin~Aspin\altaffilmark{1}}

\altaffiltext{1}{Institute for Astronomy, University of Hawaii at
  Manoa, 640 N. A'ohoku Place, Hilo, HI 96720. {\it
    caa@ifa.hawaii.edu}}

\begin{abstract} 
  We present optical and near-IR observations of the young eruptive
  variable star V1647~Orionis which illuminates McNeil's Nebula.  In
  late 2003, V1647~Ori was observed to brighten by around 5~mags to
  r'=17.7.  In early 2006 the star faded back to its quiescent
  brightness of r'$\sim$23, however, in mid-2008 it brightened yet
  again by $\sim$5~mags.  Our new observations, taken in early 2011,
  show V1647~Ori to be in an elevated photometric state with an
  optical brightness similar to the value found at the start of the
  2003 and 2008 outbursts.  Optical images taken between 2008 and 2011
  suggest that the star has remained in outburst from mid 2008 to the
  present.  H$\alpha$ and the far-red \ion{Ca}{2} triplet lines remain
  in emission with H$\alpha$ possessing a significant P~Cygni profile.
  A self-consistent study of the accretion luminosity and rate using
  data taken in 2004, 2007, 2008, and 2011, indicates that when
  bright, V1647~Ori has values of 16$\pm$2~L$_{\odot}$ and
  4$\pm$2$\times$10$^{-6}$~M$_{\odot}$~yr$^{-1}$, respectively.  We
  support the premise that the accretion luminosity and rate both
  declined by a factor 2--3 during the 5~mag fading in 2007.  However,
  a significant parts of the fading was due to either variable
  extinction or dust reformation. We discuss these new observations in
  relation to previous published data and the classification schemes
  for young eruptive variables.

\end{abstract}


\section{INTRODUCTION}
Photometric variability is often present in young low-mass stars and
can take the form of either periodic or stochastic brightness
modulations.  They are generally associated with such physical
processes as stellar rotation, changes in overlying circumstellar
obscuration, or variable mass accretion.  However, the most dramatic
variability seen is that of an eruptive nature when large amplitude
photometric changes occur over relatively short periods of time,
resulting in optical brightness increases of 100 fold or more.  Such
outbursts are thought to be an integral part of a young star's
pre-main sequence life being times when the accretion rate
dramatically increases, and hence, they are phases when a large
fraction of the star's final mass is accreted.  We refer the reader to
the review articles by Hartmann \& Kenyon (1996) and Reipurth \& Aspin
(2010) for more significant background on eruptive variables.  Two
varieties of eruptive variables have previously been found.  These are
colloquially termed FUors (Ambartsumian 1971) and EXors (Herbig 1989).
FUor outbursts are named after the first young star seen to exhibit
such a dramatic brightness increase, namely FU~Orionis (Wachmann 1954;
Herbig 1966).  In the mid-1930s, FU~Ori optically brightened by over
5~mags and has remained in this elevated state to the present day
(i.e., over 70~years).  EXors, named after the progenitor of the class
EX~Lup, are similarly energetic, however, their outbursts typically
result in a brightening of 1--4~mags.  In addition, EXor events
generally last weeks to months rather than years to decades as in
FUors.  The spectral characteristics of FUors and EXors are also
significantly different.  In outburst, FUors exhibit optical
absorption features indicative of a F--G supergiant star (Herbig 1966,
1977) and NIR absorption features typical of an M giant (Mould et al.
1978; Reipurth \& Aspin 1997).  They typically show H$\alpha$
absorption with an associated blue-shifted component creating a
P~Cygni profile (Herbig 1966, 1977).  Little to no H$\alpha$ emission
is observed.  In outburst, EXors show an optical and NIR emission
spectrum and often possess emission in the NIR CO overtone bandheads
(see for example EX~Lup, Aspin et al. 2010).  Such a dichotomy is
peculiar since, in both cases, the outbursts are thought to be driven
by enhanced mass accretion.  It remains a mystery how young star
outbursts driven by the same physical mechanism can exhibit such
different photometric and spectroscopic characteristics.

An alternative model for FUor outbursts that warrants consideration
was proposed by Larson (1980) and subsequently further investigated by
Herbig (1989), Petrov \& Herbig (1992), Herbig et al.  (2003), and
Petrov \& Herbig (2008). In this model a rapidly rotating young star
develops a bar-like deformation that causes the outer layers of the
star to heat up resulting in the observed increase in luminosity.
Although not as popular as the accretion disk model, this model does
have several attractive characteristics, none less than providing
separate outburst mechanisms for FUors and EXors.

It is clearly important to find and study in detail new examples of
both FUors and EXors.  To date, out of the tens of thousands of young
stars known, only a dozen or so of both types are known.  One recent
discovery was V1647~Ori.  This young star is located near M~78 in
Orion and was discovered by amateur astronomer Jay McNeil (McNeil
2004) when he found a previously unidentified nebula in the region.
This nebula, now known as McNeil's Nebula, is illuminated by
V1647~Ori, which appears at the apex of its monopolar geometry.  In
quiescence, V1647~Ori is a faint young star (r'$\sim$23) located near
the Herbig-Haro (HH) objects HH~22, 23, and 24 (Herbig 1974).  Since
its discovery in late 2003, soon after it had brightened by
$\sim$5~mags, it was the subject of numerous observational studies at
wavelengths spanning the electromagnetic spectrum from X-rays to
radio.  Rather than explicitly mention each of the 50+ papers
published from 2004 to the present day, we refer the reader to Aspin
\& Reipurth (2009) for both a more comprehensive list and a discussion
of the optical photometric and spectroscopic variability during the
period 2003--2006.

Soon after discovery, V1647~Ori exhibited a strong emission spectrum
in both the optical and NIR.  This eruption (henceforth referred to as
the first outburst) lasted around 2~years, after which the star faded
back to its pre-outburst quiescent brightness.  All indications were
that V1647~Ori was an example of the EXor class of eruptive variables
especially in light of the presence of a previous outburst in 1966
(Aspin et al.  2006).  However, it was subsequently shown that at high
spectral resolution, the source possessed unique 2~$\mu$m spectral
structure only found in FUors (Aspin et al. 2008).  It was somewhat
surprising that in 2008 V1647~Ori again optically brightened by
$\sim$5~mags (Kun 2008; Aspin et al. 2009).  Observations taken during
the new eruption (below referred to as the second outburst) showed
almost identical features to the earlier event (Aspin et al. 2009),
from a similarly illuminated McNeil's Nebula to an optical emission
spectrum with a P~Cygni profile H$\alpha$, and CO overtone bandhead
emission.  In the period 2008 to 2010, it is unclear what changes
occurred in the optical and NIR brightness of V1647~Ori. Few
observations of the source were reported during this period, however,
Teets et al. (2011) found that throughout the period 2002 to 2009, the
X-ray flux from V1647~Ori was well correlated with its optical and NIR
brightness. Additionally, in the period 2008 September to 2009 April,
its X-ray flux varied on short-timescales between 2 and
8$\times$10$^{30}$~ergs~s$^{-1}$ \footnotemark
\footnotetext{Pre-outburst, V1647~Ori had an X-ray luminosity of
  $\sim$3$\times$10$^{29}$~ergs~s$^{-1}$ while during the 2004
  outburst it rose to $\sim$10$^{31}$~ergs~s$^{-1}$ (Kastner et al.
  2004).}.  This result implies that V1647~Ori was highly active
during the above period and suggests that it was likely bright at
optical and NIR wavelengths.

In this paper, we detail the status of the source in the period 2011
February to April, finding that it has remained in an highly active
eruptive state.

\section{OBSERVATIONS \& DATA REDUCTION}
The complete observing log for the V1647~Ori observations presented
below is shown in Table~\ref{obslog}.  Observations were acquired on
three different telescopes, all located on Mauna Kea, Hawai'i.  The
``Fredrick C.  Gillett'' Gemini-North 8m telescope was used on five
different occasions to obtain optical and NIR imaging and spectroscopy
(under queue observing programs GN-2010A-Q-10 and GN-2011A-Q-37).
Additional NIR spectroscopy was obtained in engineering time on the
NASA IRTF 3m telescope.  Optical images of V1647~Ori were also
obtained at four different epochs on the University of Hawai'i 2.2m
telescope (from 2008 to 2009).  Below we detail the observations
acquired and briefly comment on their subsequent reduction and
analysis.

\subsection{Optical Imaging}
The Gemini-North instrument used for optical imaging was GMOS (Davies
et al.  1997; Hook et al.  2004).  Data were acquired on UT 2011
February 2.  Photometric calibration was achieved using the field star
sequence defined in Aspin \& Reipurth (2009).  We used the SDSS r'
filter with a 0$\farcs$14 pixel scale, and a total exposure time of
60s.  All images were reduced using the Gemini GMOS {\it iraf} package
(v1.10).  Photometry was extracted from the images using the Starlink
GAIA package.  Similar imaging observations were obtained with
Gemini-North/GMOS in 2004 and 2007 (Aspin \& Reipurth 2009).

The University of Hawai'i 2.2m telescope and the thinned back-side
illuminated Tektronix 2048$\times$2048 optical CCD camera, Tek, were
used to obtain optical imaging on UT 2008 August 31 and October 9,
and 2009 January 23 and September 11.  At all three epochs, a
Johnson-Cousins R$_C$ filter was used with a detector-defined pixel
scale of 0$\farcs$21. The exposure time in each case was 60s.
Photometric calibration again used the field star sequence mentioned
above.  Transformations to the SDSS r' filter system was performed as
outlined in Aspin \& Reipurth (2009).  The images were reduced using
the Starlink package CCDPACK in a standard manner. The Starlink GAIA
package was again used to perform photometry extraction.

\subsection{Optical Spectroscopy}
As in the case of the Gemini-North imaging above, the instrument used
for optical spectroscopy was GMOS.  Data were acquired on UT 2011
February 2 using the B600 grating with a central wavelength 7500~\AA.
A 0$\farcs$75 wide long-slit was used resulting in a resolving power,
{\it R}, of $\sim$1200 (0.45~\AA~pixel$^{-1}$).  This value of {\it R}
gives a full-width half maximum (FWHM) of unresolved lines of
$\sim$130~km~s$^{-1}$.  The total on-source exposure time for the
spectroscopy was 300s.  Identical observations of the
spectrophotometric standard star G191-B2B were also taken to allow
flux calibration and sensitivity function definition.  All spectra
were reduced using the Gemini GMOS {\it iraf} package (v1.10).
Feature extraction was performed using the {\it iraf} task {\bf
  splot}.

\subsection{Near-IR Imaging}
Our NIR imaging observations were taken using the Gemini-North
facility NIR camera, NIRI (Hodapp et al. 2003).  Data were acquired on
UT 2011 February 6 using the f/32 camera due to the intrinsic
brightness of the source (thus allowing shorter exposure times) in
standard ``Mauna Kea'' J, H, K', and L' filters.  For each filter,
four dithered images were taken to allow for both sky subtraction and
uncertainty estimation.  Total exposure times of 20s were used for J,
H, and K', and 48s for L'.  Similar observations of the NIR faint
standard star FS~150 were taken for flux calibration purposes.  The
images were reduced using the Gemini NIRI {\it iraf} package (v1.10).
Extraction of photometry from the reduced images was performed using
the Starlink GAIA program.  Flux from V1647~Ori was extracted using
5$''$ diameter circular apertures with sky values determined from
regions of blank sky close to the source.  Similar analysis was
performed on the standard star.  This technique gave us the
star+nebula photometry within the aperture used and it is this we
present below.

\subsection{Near-IR Spectroscopy}
Our first of two epochs of NIR spectroscopic observations were
acquired using the Gemini-North facility integral-field unit (IFU)
spectrograph, NIFS (McGregor et al.  2003) using the J, H, and K-band
gratings.  Data were obtained on UT 2011 February 15 using total
exposure times of 120s per waveband, resulting in spectra with a
spectral resolution of {\it R}$\sim$5000.  Sky observations, taken
using an 'ABA' offset sequence (where A is on-source and B is sky),
were also acquired to facilitate accurate sky subtraction.  Similar
observations of the A0~V star HIP~103222 were taken to allow the
removal of telluric features from the target spectra.  The data were
reduced using the Gemini NIFS {\it iraf} package (v1.10).  The final
spectra of the target and telluric standard are the sum of the IFU
pixels lying within an 0$\farcs$5 diameter software aperture centered
on the targets.  Telluric correction was performed using the IDL SpeX
procedure {\bf xtellcor\_general} (Vacca et al. 2003;Cushing, Vacca,
\& Rayner 2004).  Conditions were clear and the seeing stable during
the target and standard observations and hence, since this is IFU
data, we consider the flux calibration good to the 5\% level.

Our second epoch NIR spectroscopy was acquired on the NASA IRTF
telescope on UT 2011 April 19 using the facility NIR spectrograph SpeX
(Rayner et al. 2003).  The cross-dispersed (XD) mode was used and
observations were acquired using the short--XD settings resulting in a
spectral resolution of {\it R}$\sim$1500.  The data were reduced using
the SpeX IDL package (Cushing, Vacca, \& Rayner 2004). Telluric
correction and flux calibration were performed using the A0~V telluric
standard star HD~35656. Conditions were again clear and the seeing
stable during the target and standard observations and we consider the
flux calibration good to the 10\% level.

\section{RESULTS}
\subsection{The Recent Photometric Behavior of V1647~Ori}
V1647~Ori was originally observed to go into outburst in the fall of
2003 (Brice\~no et al. 2004).  This phase lasted about two years and
by the fall 2005 had faded back to its pre-outburst quiescent
brightness (r'$\sim$23, Aspin Beck, \& Reipurth 2008).  This behavior
was very similar to the previously documented outburst of the source
in 1966 (Aspin et al.  2006).  However, to the surprise of the
variable star community, in the summer months of 2008 (while Orion was
behind the Sun) V1647~Ori yet again flared to a similar outburst
brightness (r'$\sim$17).  Between 2008 and 2011 the photometric
behavior of V1647~Ori remains unclear with little to no new data being
published.  The AAVSO database of amateur astronomer observations
contains a few V-band and ``clear'' observations of the source from
2008 September to 2011 February, however, they are somewhat disparate
in nature (some observations within a few days of each other have
$\Delta$V of over 4~mag) and we cannot consider them a truly reliable
source for the recent photometric history of the young star.  In
Table~\ref{optphot} we have compiled several observations of V1647~Ori
from data acquired by cooperative observers on the University of
Hawai'i 2.2m telescope.  Although sporadic in nature, these R-band
observations suggest that between the second outburst of V1647~Ori in
mid-2008 and the early months of 2011, V1647~Ori remained in an
elevated photometric state with an optical r' brightness between 16.9
and 17.8.  Even though the gaps in our photometry are relatively large
(ranging from 1 to 9 months), we believe that the temporal range,
sampling frequency, and random nature of the observations all support
the hypothesis that from 2008 August to 2011 February (i.e., around 32
months), V1647~Ori remained in outburst.

\begin{figure}[t] 
\begin{center}
\includegraphics*[angle=0,scale=1.3]{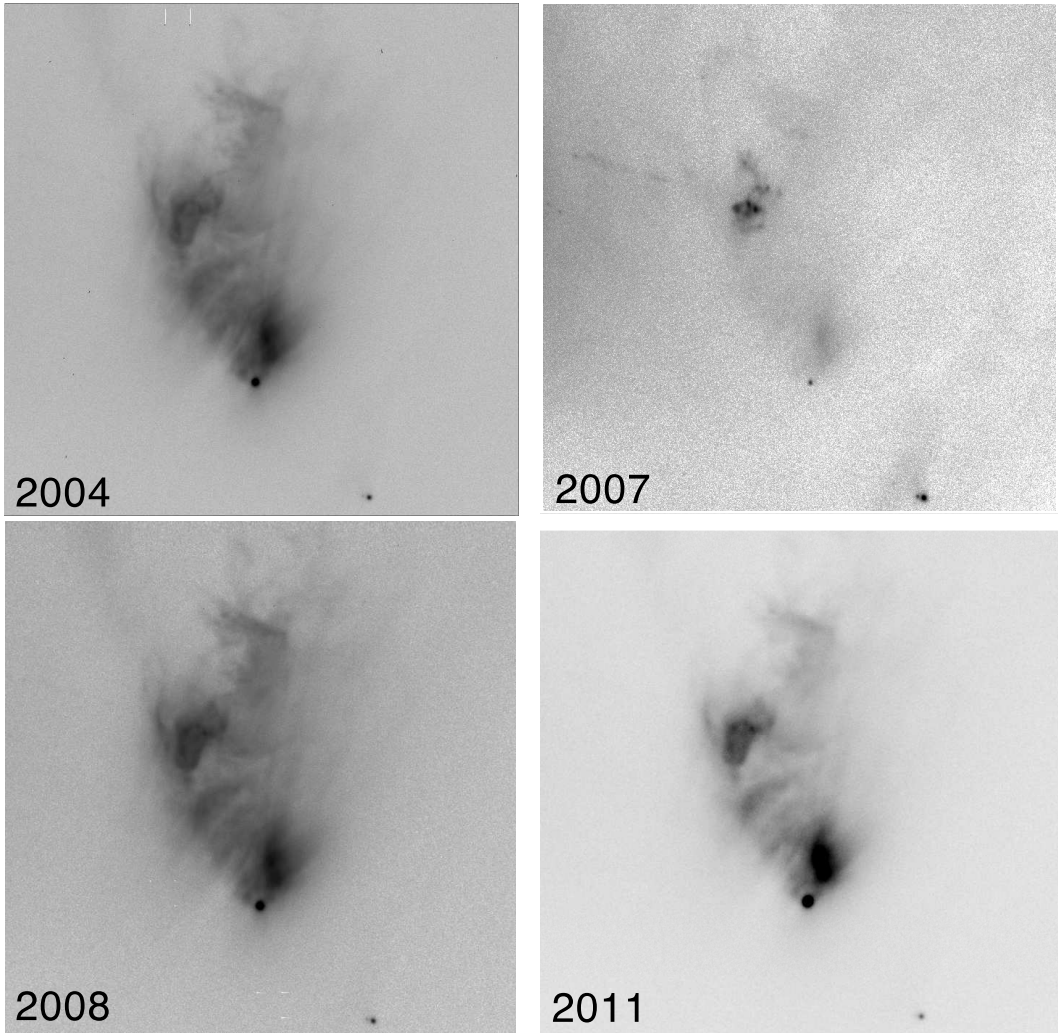} 
\caption{Optical r' images of McNeil's Nebula and V1647~Ori 
from four different epochs.  Each image is 100$''$ in size.
North is up, East to the left in all panels.
Top-left from February 2004, a few months after the start of the 2003
outburst.  Top-right from 2007, a year after the source had faded back
to its quiescent brightness level.  In this image the HH~22 complex is visible
north and slightly east of V1647~Ori.  Bottom-left 2008, soon after V1647~Ori 
had brightened for a second time.  Bottom-right 2011, taken in early 
February.  Note the similarity in structure in the 2004, 2008, and 2011
images. For comparison, the faint source to the south-west of V1647~Ori 
has an r' magnitude of 21.00$\pm$0.04 (Aspin \& Reipurth 2009).
\label{optims}}

\end{center}
\end{figure}

Figure~\ref{optims} shows r' band images of V1647~Ori and McNeil's
Nebula from four different epochs ranging from early 2004 to early
2011.  The data from 2004 (first outburst, top-left), 2007 (quiescent
phase between outbursts, top-right), 2008 (second outburst,
bottom-left) and 2011 (bottom-right) all appear morphologically
similar with all nebulous features seen in the original outburst image
from 2004 repeating in 2008 and 2011.  The 2007 image, showing the
source after the first outburst had faded (hence McNeil's Nebula is
only faintly visible), shows well the HH~22 object/jet which is
present in all images but difficult to see due to the presence of the
bright reflection nebulosity.  The nature of the structure in the
nebula is interesting in that over a period of $\sim$7~years, the
physical material which is reflecting the light from V1647~Ori appears
static.  The distance from V1647~Ori to the northern almost horizontal
section of nebulosity is 55$''$ (directly north, sloping down from
east to west).  Assuming a distance of 450~pc and a nebula axis
inclination with respect to the plane of the sky of
29$\pm$14$^{\circ}$ (Acosta-Pulido et al.  2007), this corresponds to
0.14~pc (and hence the light travel time between the two locations is
$\sim$164~days).  If the nebulous structures are located on the walls
of an evacuated cavity (created by an earlier molecular outflow as
generally assumed) and are moving at typical CO flow values of
10~km~s$^{-1}$, then even in 7 years the expected motion of the
material along the cavity walls would be only $\sim$15~AU or, at the
above distance, 0$\farcs$04.  This would clearly not be detected in
our natural seeing images. 

\subsection{NIR Color Variations}
The compilation of NIR photometry of V1647~Ori in Table~\ref{nirphot}
shows observations from five epochs ranging from 2004 March to 2011
April.  The purpose of this Table is to show the relative consistency
in photometry from the time of the original outburst (2004 March) to
soon after the second outburst (2008 September) and onto the current
observations (2011 February and April).  Data from the quiescent phase
between the first and second outburst are also shown (2007 February).
Clearly, V1647~Ori has maintained its optical {\it and} NIR outburst
brightness and colors subsequent to its second outburst.

\begin{figure}[t] 
\begin{center}
\includegraphics*[angle=0,scale=0.7]{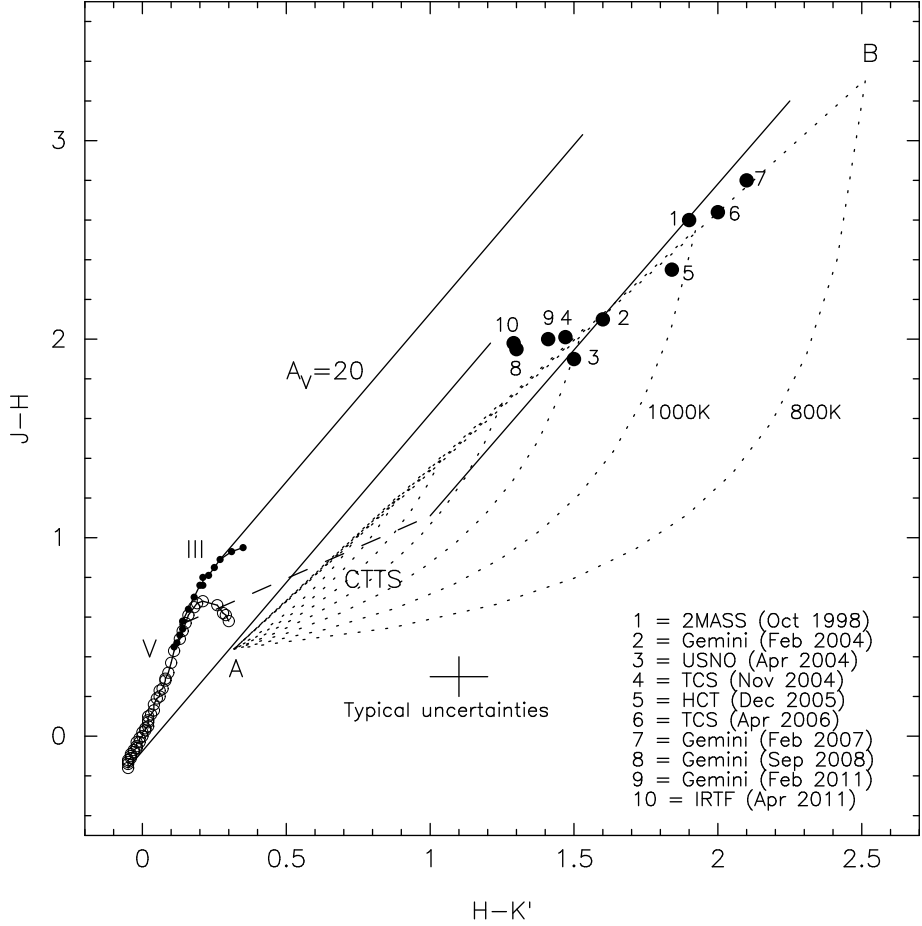} 
\caption{NIR J--H vs. H--K' color--color diagram showing the location
  of V1647~Ori at ten different epochs including the data presented
  here (points \#9 and \#10).  Typical observational uncertainties are
  shown as a cross at J--H=0.3, H--K'=1.1.  The zero-age main sequence
  (open circles, labeled V) and giant branch (small filled circles
  labeled III) are shown and have Rieke \& Lebofsky (1985) reddening
  vectors (solid lines) extending from them for A$_V$=20~mags.  The
  Meyer et al. (1997) locus of classical and weak-line T~Tauri stars
  is also shown (dashed line, labeled CTTS).  From the larger H--K' end
  of this locus extends a reddening vector also for A$_V$=20~mags
  (solid line).  The dotted lines show the loci of varying
  contributions from {\it i)} a black-body of a 'stellar' temperature
  of 4000~K and {\it ii)} black-body dust temperatures of 800 to
  2000~K in steps of 200~K.  Along the loci the contributions vary
  from from 100\% stellar (close to the ZAMS) to 100\% dust.  The
  dotted line A--B indicates the loci of pure dust emission.  Point
  \#1 is from 2MASS catalog.  Point \#2 is from Reipurth \& Aspin
  (2004).  Point \#3 is from the United States Naval Observatory
  telescope as presented in McGehee (2004). Points \#4 and \#6 are
  from the Telescopio Carlos Sanchez (TCS) at the Teide Observatory
  presented by Acosta-Pulido et al. (2007).  Point \#5 is from the
  Himalayan Chandra Telescope (HCT) presented in Ojha et al. (2007).
  Point \#7 is from Gemini presented in Aspin, Beck, \& Reipurth
  (2008).  Point\#8 is from Gemini presented in Aspin et al. (2009).
  Points \#9 and \#10 are from this work.
\label{jhkcc}}

\end{center}
\end{figure}

\begin{figure}[t] 
\begin{center}
\includegraphics*[angle=0,scale=0.7]{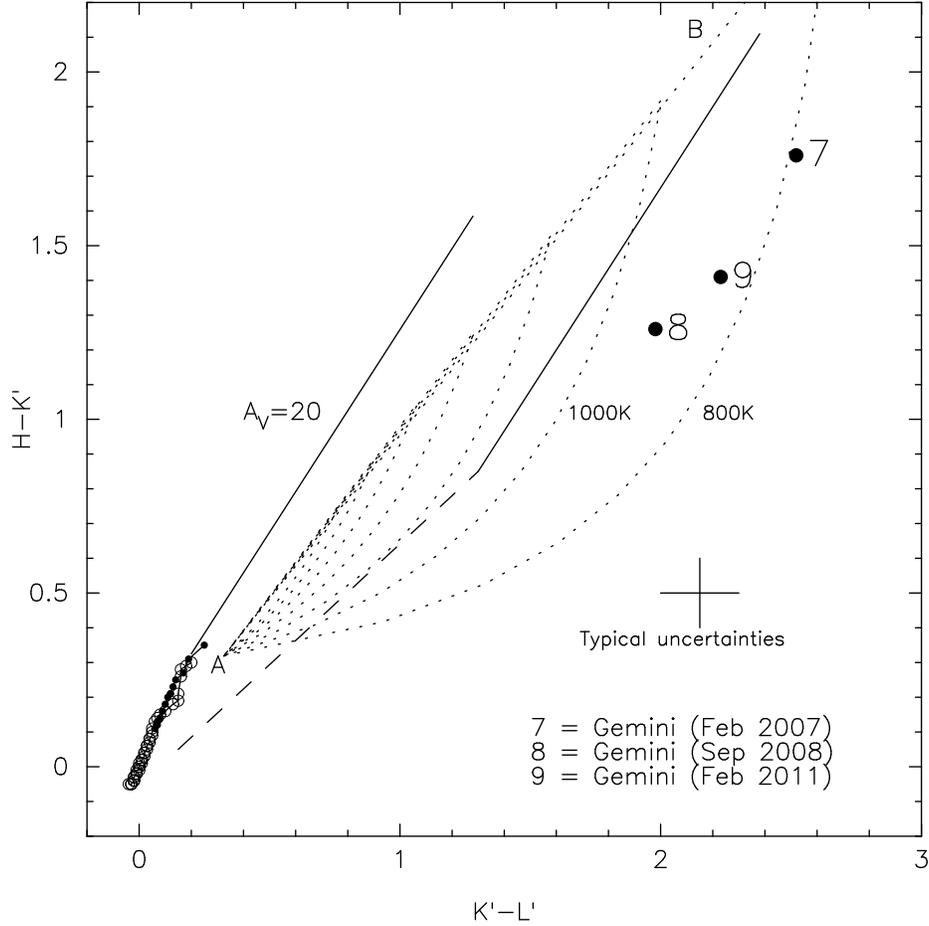} 
\caption{NIR H--K' vs. K'--L' color--color diagram showing the location
  of V1647~Ori in February 2011 (\#9). The colors of V1647~Ori from
  February 2007, and September 2008 are also plotted (point \#7, and
  \#8, respectively).  The point numbering is the same as in
  Figure~\ref{jhkcc}. Typical observational uncertainties are shown as
  a cross at H--K'=0.5, K'--L'=2.0.  The zero-age main sequence (open
  circles) and giant branch (small filled circles) are shown. Rieke \&
  Lebofsky (1985) reddening vectors (solid lines) for A$_V$=20~mags
  extend from both the extreme of the zero-age main sequence/giant
  branch and the end of the Meyer et al. (1997) locus of classical and
  weak-line T~Tauri stars (dashed line).  The dotted lines show the
  loci of varying contributions from {\it i)} a black-body 'stellar'
  temperature of 4000~K and {\it ii)} black-body dust temperatures of
  800 to 2000~K in steps of 200~K.  Along the loci the contributions
  vary from from 100\% stellar (close to the ZAMS) to 100\% dust.  The
  dotted line A to B indicates the loci of pure black-body dust
  emission.
\label{hklcc}}

\end{center}
\end{figure}

The NIR J--H, H--K', and K'-L' colors of the sources, and their
variability, are shown in Figures~\ref{jhkcc} and \ref{hklcc}.  The
J--H vs. H--K' color--color (henceforth JHK' c--c) diagram
(Figure~\ref{jhkcc}) shows the location of V1647~Ori at 10 different
epochs (including the five shown in Table~\ref{nirphot}).  These
points, labeled \#1 through \#10, are in chronological order and show
significant color variability during the period 1998 (2MASS) to 2011.
Other than points \#8--10, the values were compiled from other sources
(listed in the figure caption).  Aspin, Beck, \& Reipurth (2008)
interpreted the color variation seen in Figure~\ref{jhkcc} (points
\#1--7) as due to changes in overlying extinction caused by either the
sublimation of dust in the outburst, or denser material moving into
the line-of-sight e.g. similar to what occurred in KH~15D (see
Kusakabe et al. 2005 and references therein) since the variations were
directly along standard reddening vectors.  The additional photometric
points (\#8--10) are all consistent with the values of the JHK' colors
at the time of the first outburst in 2004 March.  Point \#8 is from
soon after the second outburst phase.  Points \#9 and \#10 are from
2011 February and April, respectively.  This behavior is consistent
with the optical brightness of V1647~Ori at these times; the infrared
colors are smaller during the outburst while they are larger colors in
quiescence.

In Figure~\ref{jhkcc} we have plotted loci of varying contributions of
a stellar photosphere at a black-body temperature of 4000~K and dust
emission from black-body temperatures of 800--2000~K (dotted lines).
At position ``A'' the contribution is 100\% from the stellar
photosphere while at position ``B'' it is 100\% from the heated dust.
The simple model used generated black-body fluxes using the IDL
routine {\bf PLANCK} which takes as input the required wavelength and
temperature.  We generated fluxes at the central wavelengths of the
NIR filters for both the stellar temperature and the heated dust
temperature and then divided both by the corresponding values from an
A0 star (at 10,000~K).  These scaled fluxes were then normalized to
unity at the middle wavelength (i.e. the H-band for J, H, and K
filters) and the results used to calculate the NIR colors in the
standard manner.  Although the generated fluxes are not integrated
over specific filter profiles, we consider that the values at the
central wavelengths of the filters are sufficiently accurate for our
needs.  The contributions to the derived colors from the stellar
photospheric and heated dust fluxes vary from 0 to 1 and 1 to 0,
respectively.  This results in the (dotted) lines shown in
Figure~\ref{jhkcc}.

One interesting fact is that the line joining points A and B is close
to being parallel to the reddening vector from the extreme of the
Meyer et al. (1997) classical T~Tauri star locus (dotted line).  One
could therefore interpret the change in colors of V1647~Ori as being
the result of changing black-body dust temperature from $\sim$900~K at
point \#7 to $\sim$1200~K close to the group of points centered on
point \#9.  However, this interpretation would assume that all the
emission from the source originates as black-body dust emission with
no stellar photospheric contribution.  This seems unreasonable for
three reasons {\it i)} the source is optically bright, {\it ii)}
Aspin, Beck, \& Reipurth (2008) found photospheric absorption features
in the NIR spectrum of V1647~Ori during quiescence (in 2007), and {\it
  iii)} the low inclination of the outflow cavity
($\sim$29$\pm$14$^{\circ}$ with respect to the plane of the sky) and
the lack of a southern counterpart outflow lobe suggests we are not
looking through dust dominated material (i.e., through the disk) along
our line-of-sight to the circumstellar regions.  We therefore consider
that our new data supports the interpretation of the change in JHK'
colors as due to extinction changes resulting from the outburst event
as proposed earlier by Aspin \& Reipurth (2009).

In Figure~\ref{hklcc} we show the H--K' vs K'--L' (henceforth HK'L')
color variations for V1647~Ori.  Three points from different epochs
are plotted (\#7--9), specifically, 2007 February, 2008 September, and
2011 February\footnotemark \footnotetext{The point numbering
  corresponds to those in Figure~\ref{jhkcc}.}.  These three points
lie parallel to the reddening vectors consistent with the suggestion
that the changes observed are due to variations in overlying
extinction.  Point \#7 has the largest colors in both the JHK' and
HK'L' c--c diagrams and was taken in 2007 February, between the first
and second outbursts.  Point \#8 has the smallest colors in both
diagrams and at that time (2008 September), V1647~Ori was in a bright
phase just after its second outburst.  Point \#9 represents the latest
observation (2011 February) and, within the associated uncertainties,
is very similar to the 2008 elevated state value (\#8).  The A$_V$
difference between point \#8 and \#9 amounts to only a few mags in
both c--c diagrams (1.5 in JHK' and 3 in HK'L').  The A$_V$
differences between points \#7 and \#8 is $\sim$8~mags in both
diagrams.

\subsection{Optical Spectroscopic Features}
\begin{figure}[t] 
\begin{center}
\includegraphics*[angle=270,scale=0.65]{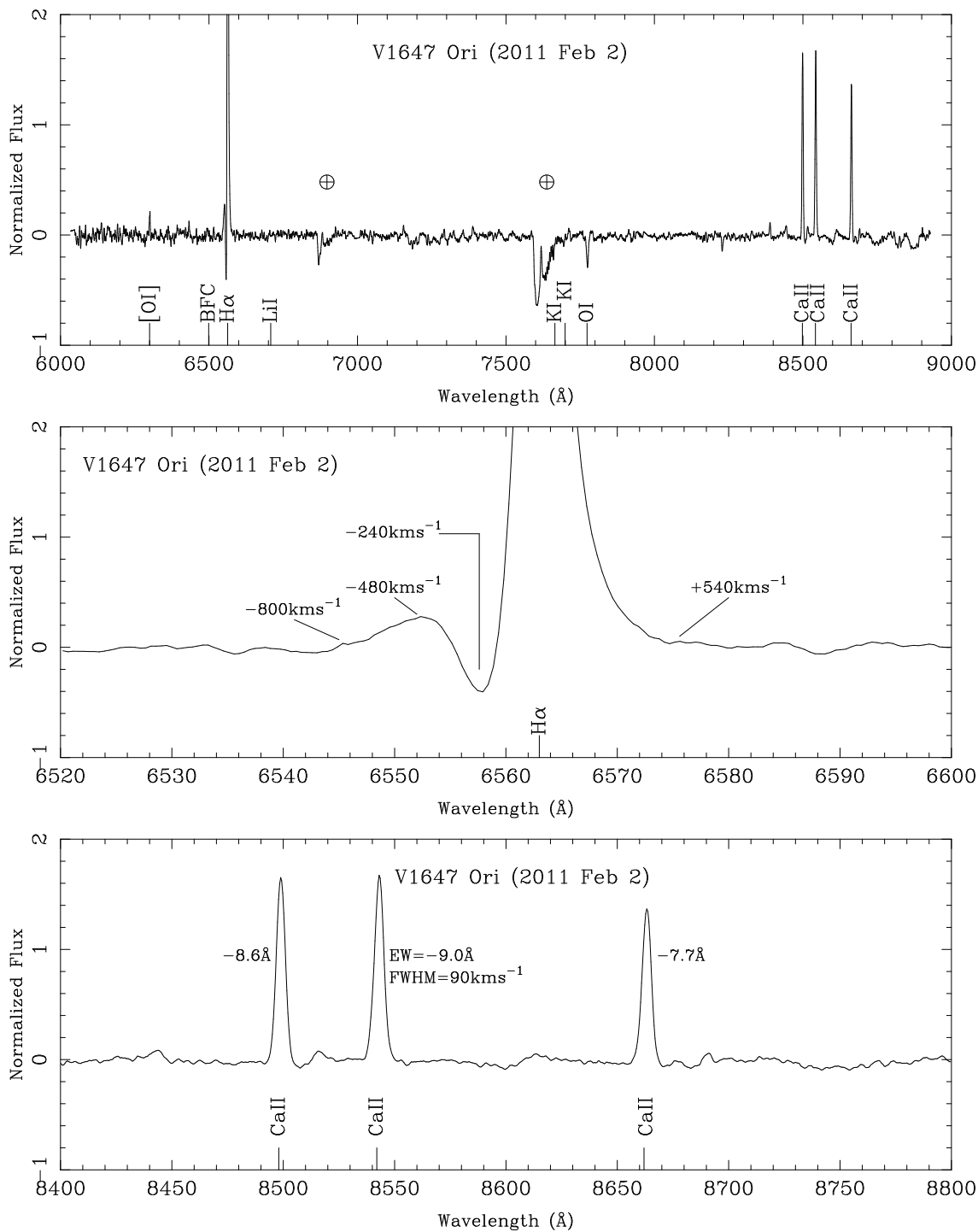} 
\caption{Optical spectrum of V1647~Ori from UT 2011 Feb 2.  This
  Gemini/GMOS spectrum covers the wavelength range 6000--9000~\AA\ and
  was taken using the B600 grating and a 0$\farcs$75 wide long-slit
  resulting in a resolving power of R$\sim$1500.  Top is the full continuum
  subtracted spectrum showing H$\alpha$ in emission with a P~Cygni
  profile.  Also in emission are the far-red \ion{Ca}{2} triplet
  lines.  $[$\ion{O}{1}$]$ at 6300~\AA\ is in emission while
  \ion{O}{1} at 7775~\AA\ is in absorption.  Middle is an expanded
  view of the region surrounding H$\alpha$ showing the blue-shifted
  absorption.  Bottom is the region around the \ion{Ca}{2} triplet
  lines.
  \label{optspec}}

\end{center}
\end{figure}

During the two recent outburst phases (i.e., 2004 March and 2008
September), V1647~Ori typically showed an optical emission spectrum
which included a strong H$\alpha$ line with associated sub-continuum
blueshifted absorption (i.e., a P~Cygni profile) and strong far-red
\ion{Ca}{2} triplet emission.  Examples of these spectra can be found
in Aspin \& Reipurth (2009).  In its faint phase between the above
outbursts (i.e., in 2007 February), the source also exhibited an
optical emission spectrum but with significantly weaker H$\alpha$
emission (with no sub-continuum blueshifted absorption) plus, as in
outburst, \ion{Ca}{2} triplet line emission.  Our 2011 February
optical spectrum of V1647~Ori is shown in Figure~\ref{optspec}.  The
line fluxes extracted from this spectrum are shown in
Table~\ref{optfluxes}.  The top panel shows the complete GMOS spectrum
covering the wavelength range 6000--9000~\AA.  This spectrum has been
continuum subtracted to enhance the visibility of the spectral
features present\footnotemark \footnotetext{The optical continuum
  simply rises to the red through the above wavelength region.}.  In
this spectrum, H$\alpha$ is, as expected, in emission and, like the
aforementioned outburst phase spectra, shows a strong sub-continuum
blueshifted absorption component.

Closer views of the region surrounding H$\alpha$ and \ion{Ca}{2}
triplet lines are shown in the middle and bottom panels of
Figure~\ref{optspec}.  The wings of H$\alpha$ extend out to --800 and
+540~km~s$^{-1}$, on the blue and red side, respectively, and the
absorption minimum occurs at --240~km~s$^{-1}$.  The \ion{Ca}{2}
triplet lines are also in emission and it seems they are marginally
resolved with deconvolved FWHM values of $\sim$90~km~s$^{-1}$.  The
ratios of the EW values for the three lines (i.e., --8.6, --9, and
--7.7~\AA) are 1.9:2:1.7.  This places V1647~Ori in the highly
optically thick region of Figure~3 from Herbig \& Soderblom (1980) and
Figure~8 of Hamann \& Persson (1992).

\begin{figure}[t] 
\begin{center}
\includegraphics*[angle=0,scale=0.8]{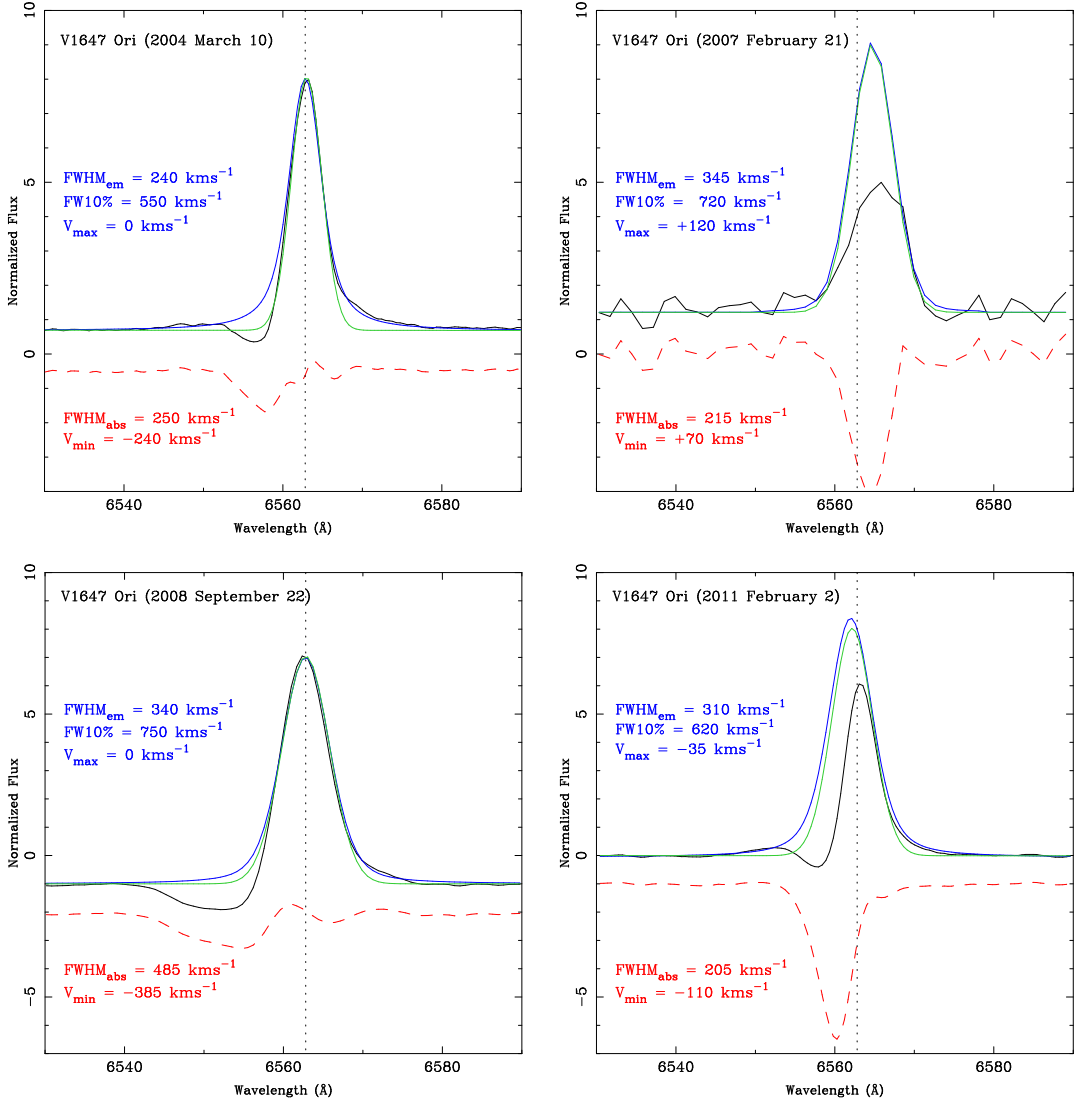} 
\caption{H$\alpha$ emission profiles (black) from four different epochs 
overlaid with a model profile (blue). The model parameters were 
determined by fitting the red wing of the emission.  The extended 
wings on the line profiles from 2004, 2008, and 2011 are best-fit 
using a Voigt profile.  The best-fit Gaussian is also shown (green).  
The dashed (red) line is the different between the observed
and best-fit model profile i.e., the associated absorption component.  The 
parameters of the model emission (blue) and absorption (red) components 
for all profiles are given for each epoch.
\label{haprof}}

\end{center}
\end{figure}

Since H$\alpha$ is clearly a composite of emission and absorption
components, we have fit an emission profile to the red wing of the
H$\alpha$ emission (the wing not effected by the blueshifted
absorption component) and attempt to determine the underlying
structure of the emission.  Figure~\ref{haprof} shows the result of
this analysis together with similar analysis for H$\alpha$ in 2004,
2007, and 2008.  Due to the high velocity wings of the emission, the
2004, 2008, and 2011 spectra are best fit with a composite Voigt
profile (Gaussian + Lorentzian).  For the 2007 spectrum, a simple
Gaussian suffices.  In all four panels, the black line is the observed
H$\alpha$ profile, the blue line is the best-fit Voigt emission
profile, the green line is the best-fit Gaussian, and the red dashed
line is the difference (observed--model).  In 2004, 2008, and 2011,
the model emission component peak occurs very close to the nominal
H$\alpha$ wavelength of 6562.8~\AA.  In 2007 (between the outbursts),
the peak is redshifted by $\sim$120~km~s$^{-1}$.  Similarly, in 2004,
2008, and 2011, the absorption component minimum, resulting from
(observed--model) emission profiles, is significantly blueshifted
(--240, --385, and --110~km~s$^{-1}$, respectively) while in 2007, the
absorption minimum is slightly redshifted (+70~km~s$^{-1}$).  This
implies that in 2004, 2008, and early 2011, there was both significant
mass accretion from disk to stellar surface and significant mass
outflow from a stellar/disk wind.  In 2007, the redshifted nature of
both emission and absorption suggests that both the H$\alpha$ emitting
and absorbing gas is associated with infalling material.

\begin{figure}[t] 
\begin{center}
\includegraphics*[angle=0,scale=0.6]{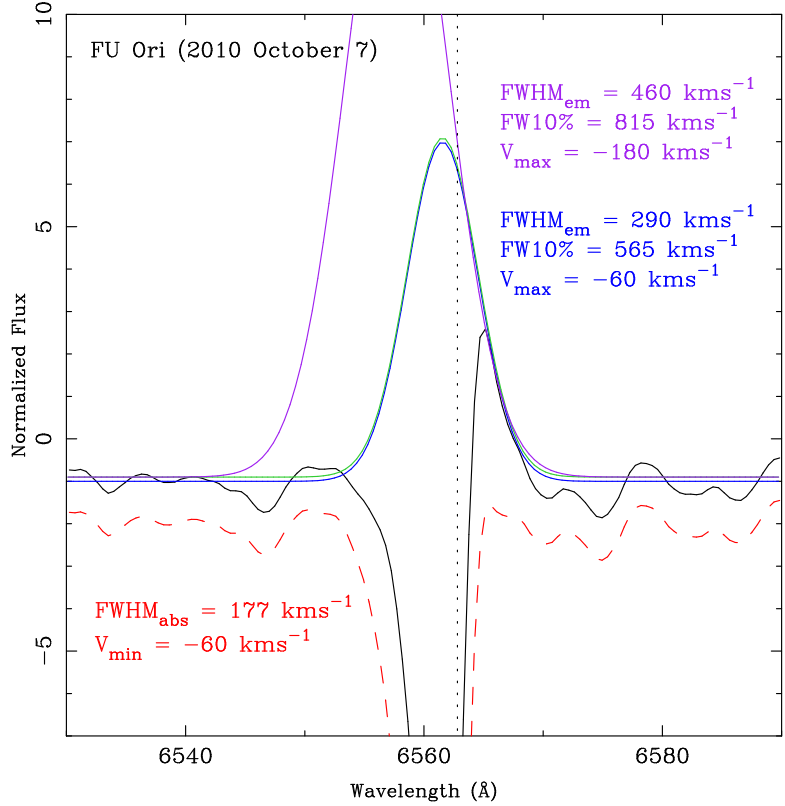} 
\caption{H$\alpha$ emission profile (black) for FU~Orionis
overlaid with a model profile (blue). The model parameters were 
determined by fitting the red wing of the emission.  The best-fit 
Gaussian is also shown (green).  The dashed (red) line is the 
different between the observed and best-fit model profile i.e., 
the associated absorption component.  The parameters of the model 
emission (blue) and absorption (red) components are given.  The 
purple curve is the required profile to give an \.M$_{acc}$ value 
of 1$\times$10$^{-5}$~M$_{\odot}$~yr$^{-1}$ using the Natta et al. 
(2004) relationship between H$\alpha$ FW10\% and \.M$_{acc}$.
\label{haprof2}}

\end{center}
\end{figure}

For comparison with the above V1647~Ori profiles, in
Figure~\ref{haprof2} we show a similar analysis of the H$\alpha$
profile of FU~Orionis from Gemini/GMOS B600 grating
spectrum\footnotemark \footnotetext{i.e., the same setup as used for
  the acquisition of the V1647~Ori GMOS spectrum.} taken on UT 2010
October 7.  We note that here, both emission and absorption components
are slightly blueshifted (by --60~km~s$^{-1}$).  The emission
component (blue/green line) is reproduced with a simple Gaussian with
a FWHM of 290~km~s$^{-1}$ and full-width 10\% intensity, FW10\%, of
580~km~s$^{-1}$.  The absorption component (red dotted line) has a
FWHM=177~km~s$^{-1}$, and is skewed with an extended blue-wing.  The
fact that the peak emission and minimum absorption components have
very similar velocity offset results in an absorption dominated
observed profile with only weak redshifted emission.  The purple line
will be discussed further below.

Table~\ref{ha} details the velocity structure present in H$\alpha$
during the two outbursts (2004 and 2008), in the quiescent phase
between them (2007), and in our latest spectrum (2011).  Also included
is the same information for FU~Ori.  The velocity information for
V1647~Ori compares well with that from FU~Ori.  Comparing these values
suggests that in 2011, V1647~Ori was very similar in characteristics
to when it was observed in 2004 and 2008.  This, coupled with the
suggestions that the source has remained bright since the second
outburst implies that V1647~Ori has remained in an elevated eruptive
state for at least the last 3~yrs.

\subsection{NIR Spectroscopic Features}
\begin{figure}[t] 
\begin{center}
\includegraphics*[angle=270,scale=0.55]{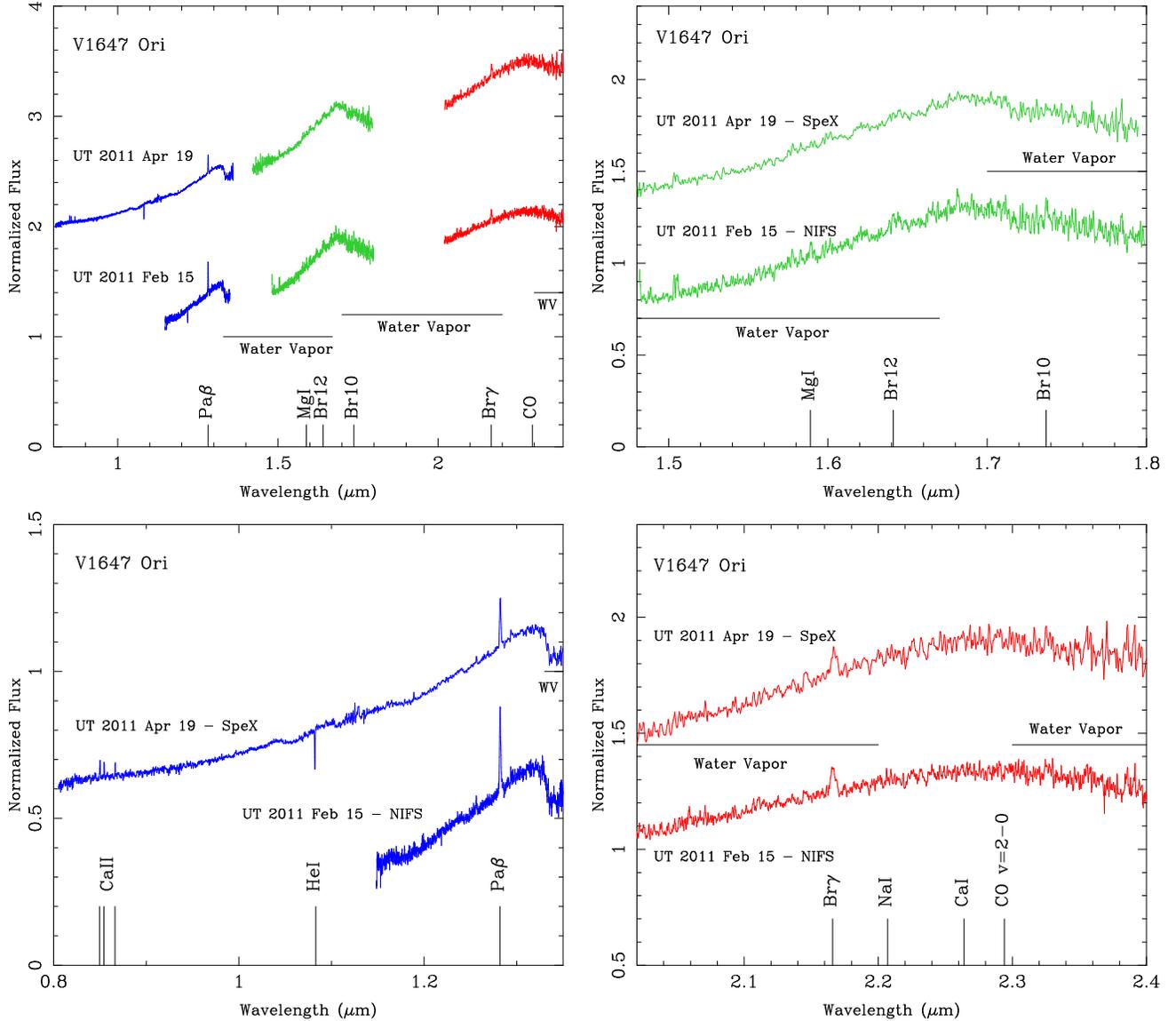} 
\caption{NIR spectra of V1647~Ori covering the wavelength region 
0.8--2.4~$\mu$m.  Spectra from two epochs are shown, specifically
UT 2011 Feb 15 and Apr 19.  The bottom of the two spectra per panel
is the Gemini/NIFS spectrum with R$\sim$5000.  The top spectra per
panel is the IRTF/SpeX spectrum with R$\sim$1500.  The top spectra 
have been shifted vertically by +1 units for clarity.
Top-left is the full wavelength range
observed.  Bottom-left, top-right, and bottom-right are closer 
views of the J-band (blue), H-band (green), and K-band (red), 
respectively.  The main features
in these spectra are strong water vapor absorption bands (indicated
by horizontal solid lines), the 
H emission lines Pa$\beta$ and Br$\gamma$, and \ion{He}{1} in absorption. 
No CO bandhead emission 
nor absorption is observed in either spectra.
\label{nirspec}}

\end{center}
\end{figure}

Our two epochs of NIR spectroscopic observations of V1647~Ori are
shown in Figure~\ref{nirspec}.  The four panels show the full
wavelength range of the data (top-left), plus individual panels for
the J (bottom-left, blue), H (top-right, green), and K (bottom-right,
red) bands.  The Gemini/NIFS spectra from UT 2011 February 15 is the
bottom of the two spectra while the IRTF/SpeX spectra from UT 2011
April 19 is at the top.  The SpeX spectrum extends to shorter
wavelengths than that from NIFS although the NIFS spectrum is of
somewhat higher spectral resolution.  The spectra from the two epochs
are very similar, both being dominated by strong water vapor
absorption bands extending over the wavelength ranges
1.3--1.65~$\mu$m, 1.7--2.2~$\mu$m, and 2.3~$\mu$m to the long
wavelength extreme of the K-band\footnotemark \footnotetext{The water
  vapor bands are intrinsic to the source and not telluric features.}.
The individual passband plots show that the spectra from the two
epochs exhibit almost identical spectral details.  The main
atomic features present are Pa$\beta$ (1.208~$\mu$m) and Br$\gamma$
(2.166~$\mu$m), both in emission.  The SpeX spectrum includes both the
\ion{He}{1} (1.082~$\mu$m) line, which is in absorption, and the
far-red \ion{Ca}{2} triplet lines, which are in emission (as they were
in our optical spectra presented above).  Line fluxes from both the
above NIR spectra are shown in Table~\ref{nirfluxes}.  In the SpeX
spectrum, the \ion{Ca}{2} triplet lines have ratios of 2.4:2:1.5 (or
in the units of Hamann \& Persson 1992, 1.2:0.7).  These values are
somewhat different from the values encountered 2.5 months earlier (see
above).  In these data, the location of V1647~Ori, in Figure~8 of
Hamann \& Persson (1992) and Figure~3 of Herbig \& Soderblom (1980),
is close to the region occupied by classical T~Tauri stars (henceforth
CTTSs) with the nearest being CW~Tau.

\begin{figure}[t] 
\begin{center}
\includegraphics*[angle=270,scale=0.8]{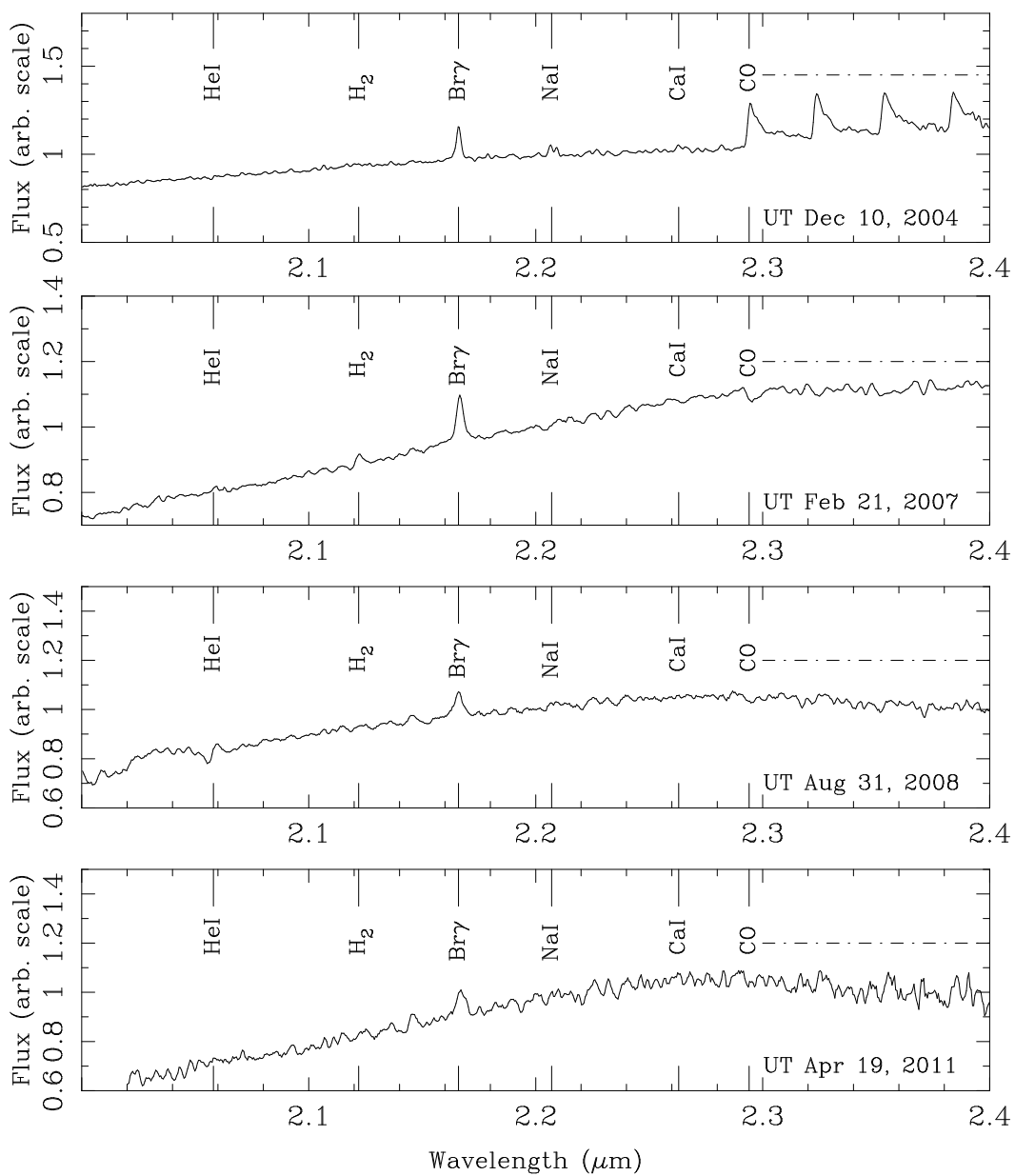} 
\caption{SpeX K-band spectra of V1647~Ori from December 2004 (top, first 
outburst), February 2007 (top-middle, between outbursts), August 2008 
(bottom-middle, second outburst), and 
April 2011 (bottom, this paper).  The main spectral features in this
region are indicated 
including the CO overtone bandheads longward of 2.294~$\mu$m (dot-dashed
line).  The spectra from 2004, 2007, and 2008 are taken from Aspin et al.
(2009).
\label{nirspec2}}

\end{center}
\end{figure}

Conspicuously absent from the K-band spectra are the CO overtone
bandheads that, when present, extend redward from 2.294~$\mu$m.  In
both K-band spectra, no CO bandheads are observed with the spectra
being featureless at those wavelengths.  This is in contrast to the
NIR spectra taken during the 2004 outburst (Reipurth \& Aspin 2004)
which showed strong CO overtone emission bands.  In 2007 (between
outbursts, Aspin, Beck, \& Reipurth 2008) and in 2008 (during the
second outburst, Aspin et al. 2009), the CO bandheads were weakly in
absorption.  Figure~\ref{nirspec2} shows an inter-comparison of the
1--2.5~$\mu$m spectral structure in 2004, 2007, 2008, and 2011.

\subsection{Accretion Luminosities and Rates}
Several methods have been presented in the literature to allow
estimation of accretion luminosities and rates from observed
quantities.  Almost all, however, rely on an accurate determination of
overlying extinction so that true dereddened fluxes can be determined.
Dahm (2008) presented a comprehensive discussion of the techniques
involved and applied them to young solar mass stars in IC~348.  He
found that estimating accretion parameters from optical and NIR
emission line fluxes (using predictions made by standard
magnetospheric accretion models, e.g. Muzerolle et al. 1998, 2001;
Kurosawa et al.  2006) and continuum excess emission determination
(Valenti et al.  1993; Gullbring et al. 1998) were in reasonable
agreement for the sample of quiescent young stars observed.  It is
unclear, however, if such an agreement can be obtained for eruptive
variables such as V1647~Ori.  The one method that does not rely on
extinction is the measurement of the FW10\% of the H$\alpha$ emission
as suggested by White \& Basri (2003) and Natta et al. (2004).  Using
equation (1) from Natta et al.  (2004) we can estimate the accretion
rate from the H$\alpha$ FW10\% value.  Below therefore, we derive
values for accretion luminosity (L$_{acc}$) and rate (\.M$_{acc}$) for
V1647~Ori in 2011 using both our NIR observations (since the
effect of reddening is significantly smaller at these wavelengths than
in the optical) and the FW10\% value for H$\alpha$ (which is not
effected by extinction).

\subsubsection{Using Pa$\beta$ and Br$\gamma$ to derive \.M$_{acc}$}
Muzerolle et al. (1998) demonstrated how dereddened Pa$\beta$ and
Br$\gamma$ emission line fluxes are correlated with mass accretion
rate in CTTSs.  The relationship determined by Muzerolle et al.
(1998) was used by Aspin, Beck, \& Reipurth (2008) and Aspin et al.
(2009) to obtain estimates of L$_{acc}$ and \.M$_{acc}$ for V1647~Ori
during the faint period between the two outbursts and soon after the
start of the second outburst.  Since both the correction to the
observed line flux (to account for overlying visual extinction), and
the stellar/disk parameter values used are of importance in such an
analysis, we here detail a self-consistent analysis using a fixed set
of stellar/disk characteristics and A$_V$ values obtained from the
JHK' c--c diagram shown in Figure~\ref{jhkcc}.  The JHK' photometry of
the source during the first outburst (epoch 2004 March, point \#2 in
the above figure), the second outburst (epoch 2008 August, point \#8),
and in early 2011 (points \#9 and \#10) all result in very similar NIR
colors.  If we dereddened their average location (along the reddening
vector shown in Figure~\ref{jhkcc}) to the CTTS locus of Meyer et al.
(1997), we obtain an A$_V$=8$\pm$2~mags.  During the faint phase
between the two outbursts (epoch 2007 February, point \#7), a similar
dereddening gives an A$_V$=19$\pm$2~mags.

It was suggested by Aspin et al. (2009) that Pa$\beta$ emission
(1.28~$\mu$m) in V1647~Ori is very likely optically thick due to the
significant deviation of the ratio Pa$\beta$/Br$\gamma$ from that
obtained for Case~B recombination theory (Hummer \& Storey 1987).
Subsequently, they only used the Br$\gamma$ emission line flux for
their estimate of L$_{acc}$ and \.M$_{acc}$.  In Table~\ref{pabbrg} we
show the results of our analysis using both Pa$\beta$ and Br$\gamma$
line fluxes from 2004 March, 2007 February, 2008 August, and 2011
February and April.  Also shown are the A$_V$ value used in
dereddening and the above line flux ratio values.  Case~B
recombination theory results in a value of P$\beta$/Br$\gamma$ of
$\sim$6 for temperatures in the range T=7500--20,000~K and electron
densities in the range n$_e$=10$^4$--10$^{10}$~cm$^{-3}$ (Hummer \&
Storey 1987, Table~6).  The only value of Pa$\beta$/Br$\gamma$ that is
close to this is from 2007 February when it had a value of $\sim$5.
This date corresponds to the faint period between the first and second
outbursts.  At all other times (i.e., early in the first and second
outburst and in early 2011), the ratio is much smaller with a typical
value of $\sim$2.  We interpret this as evidence that the Pa$\beta$
emission is optically thick during periods when the star is bright,
and close to being optically thin during its fainter periods.  This is
supported by the similarity of the derived values for L$_{acc}$ and
\.M$_{acc}$ using the Pa$\beta$ and Br$\gamma$ line fluxes in 2007
February and the significantly different values at the other three
epochs.  The conclusion we can draw from Table~\ref{pabbrg} is that
close to the start of the first and second outbursts, and in 2011, the
accretion luminosity was L$_{acc}\sim$16$\pm$2~L$_{\odot}$ and the
accretion rate was
\.M$_{acc}\sim$4$\pm$2$\times$10$^{-6}$~M$_{\odot}$~yr$^{-1}$.  In the
faint phase between the first and second outbursts, the accretion
luminosity and rate both declined by a factor $\sim$3.  Although the
statistical significance of this decline is relatively small, we
postulate that it is real since something clearly changed in
2007 to cause the 5~mag fading of V1647~Ori.  It is unlikely
that this decline is a result of an incorrect A$_V$ value since it
would require an A$_V\sim$25~mags (instead of 19$\pm$2~mags) to give
the factor 3 decline in the observed L$_{acc}$ and \.M$_{acc}$.  

\subsubsection{Using H$\alpha$ FW10\% to derive \.M$_{acc}$}
The relationship between the FW10\% width of the H$\alpha$ emission
profile and the accretion rate derived by Natta et al. (2004) relies
on the assumption that the H$\alpha$ emission is generated by the
accretion process.  Under this assumption, the relationship is linear
in nature, although the spread of values around their best-fit line is
significant.  The formula is reproduced below in Eqn (1).

\begin{equation}
  \dot{M}_{acc} = -12.89(\pm0.3) + 9.7(\pm0.7) \times 10^{-3} \times FW10\%
\end{equation}

The stars upon which the relationship was defined were a combination
of young brown dwarfs of spectral type M6--M8.5 with accretion rates
in the range 10$^{-11}$--10$^{-9}$~M$_{\odot}$~yr$^{-1}$, and young
T~Tauri stars (M$_{*}>$0.3~M$_{\odot}$) with accretion rates in the
range 10$^{-9}$--10$^{-6}$~M$_{\odot}$~yr$^{-1}$.  Over all the
sources, the range of H$\alpha$ emission FW10\% values was
200--700~km~s$^{-1}$.

The above correlation has implicit uncertainties defined by both the
errors on the observed quantities and the non-simultaneity of the
measurement of H$\alpha$ FW10\% and the independent determination of
\.M$_{acc}$.  Natta et al.  state that the derived value of
\.M$_{acc}$ should be ``used with care'' for individual objects.
Nevertheless, it is informative to apply this analysis to the data on
V1647~Ori and compare the results with those obtained from the
Br$\gamma$ line flux measurements.

Table~\ref{natta} shows the results of this analysis for V1647~Ori for
the four epochs all of which are close in time to those used to derive
\.M$_{acc}$ from the Pa$\beta$ and Br$\gamma$ fluxes (and shown in
Table~\ref{pabbrg}).  Comparing the derived values of \.M$_{acc}$ in
Table~\ref{natta} with those in Table~\ref{pabbrg}, we see significant
differences, in all cases the \.M$_{acc}$ values determined from the
H$\alpha$ FW10\% value are considerably smaller than those estimated
from the Br$\gamma$ flux.  For example, on 2004 March 9, the
dereddened Br$\gamma$ flux gives a value of
\.M$_{acc}$=5.3$\times$10$^{-6}$~M$_{\odot}$~yr$^{-1}$.  One day
later, the H$\alpha$ FW10\% gave
\.M$_{acc}$=2$\times$10$^{-8}$~M$_{\odot}$~yr$^{-1}$ some 265$\times$
smaller.  Using the relationship of Natta et al. (2004), we would
require a value of FW10\% of 785~km~s$^{-1}$ (instead of the observed
530~km~s$^{-1}$) to obtain the \.M$_{acc}$ value derived from the
Br$\gamma$ line flux.  A similar situation occurs for all four epochs
of H$\alpha$ FW10\% measurements.  In addition, the FW10\% value for
FU~Ori (580~km~s$^{-1}$) gives
\.M$_{acc}$=5$\times$10$^{-8}$~M$_{\odot}$~yr$^{-1}$ which is
significantly smaller than the expected value for a classical
FUor\footnotemark \footnotetext{A classical FUor is one whose rise
  from quiescence to outburst was documented.  Examples of classical
  FUors are FU~Ori, V1057~Cyg, V1515~Cyg, and V1735~Cyg.} of
$>$10$^{-5}$~M$_{\odot}$~yr$^{-1}$ (Hartmann \& Kenyon 1996). To
obtain an \.M$_{acc}$ value of
1$\times$10$^{-5}$~M$_{\odot}$~yr$^{-1}$ would require a FW10\% value
of 815~km~s$^{-1}$.  A best-fit profile for such an FW10\% is shown in
Figure~\ref{haprof2} (purple line).  Even if we consider the
uncertainties inherent in Eqn (1), which result in a wide range of
possible \.M$_{acc}$ values (see Table~\ref{natta}), in all cases they
do not include those values derived from the dereddened Br$\gamma$
flux.  Again using the 2004 March 10 FW10\% value as an example, we
find that the range of valid \.M$_{acc}$ values,
4$\times$10$^{-9}$--8$\times$10$^{-8}$~M$_{\odot}$~yr$^{-1}$, does not
include the one determined from the dereddened Br$\gamma$ flux,
(5.3$\times$10$^{-6}$~M$_{\odot}$~yr$^{-1}$).

We conclude, therefore, that the above derivation of \.M$_{acc}$ for
V1647~Ori using the H$\alpha$ FW10\% value as defined by Natta et al.
(2004), tends to give considerably smaller values than those
determined from dereddened Pa$\beta$ and Br$\gamma$ emission line
fluxes.  Since this also appears to be the case for the classical FUor
FU~Orionis (when compared to the definitive values of 10$^{-4}$ to
10$^{-5}$~M$_{\odot}$~yr$^{-1}$) then perhaps using the H$\alpha$
FW10\% width to estimate \.M$_{acc}$ is not valid for high accretion
rate eruptive variables such as V1647~Ori and FU~Ori.  One possibility
is that the strong star/disk wind present in FUors significantly
contributes to or modifies the H$\alpha$ emission observed (as
suggested by Lima et al.  2010).

\section{Discussion} 
From the data and analyses presented above, we support the hypothesis
that V1647~Ori has remained in an elevated eruptive state since the
time of the second outburst which occurred in mid 2008.  Although the
star likely underwent a decline in outburst activity in early
2007\footnotemark \footnotetext{We note that this decline is of
  relatively low significance with respect to the uncertainties on the
  derived values.}, its accretion luminosity and rate remained at the
upper end of the range seen in CTTSs (Hartigan, Edwards, \& Ghandour
1995; Gullbring et al. 1998).  At that time, the decline in accretion
characteristics was accompanied by a change in overlying visual
extinction, caused by either the reformation of circumstellar dust or
motion of dust into the line-of-sight.  This resulted in an increase
in extinction of over 10~mags.  At the time of the second outburst,
the enshrouding dust either re-sublimated or rotated out of the
line-of-sight, and has remained absent.

Although both the optical/NIR line emission characteristics of
V1647~Ori and the apparent decline in its optical brightness after
$\sim$2~yrs more resemble those of the shorter period EXor variables,
the source has several features in common with FUors e.g.  the
similarity of the NIR spectral absorption characteristics e.g. the
presence of strong water absorption bands and the unique spectral
structure, identified at high spectral resolution in FUors and
V1647~Ori (Aspin, Greene, \& Reipurth 2009).  However, V1647~Ori
exhibits several peculiarities that are not seen in the classical
FUors.  Specifically,

\begin{itemize}

\item The characteristic optical absorption spectrum of a F--G type
  supergiant, which all four classical FUors (FU~Ori, V1057~Cyg, and
  V1515~Cyg) exhibit, has not been seen.  V1647~Ori has instead shown
  few absorption lines and strong emission features such as H$\alpha$
  and the far-red Ca~II triplet (see Figure~\ref{optspec}).

\item The four classical FUors exhibit deep and highly variable
  blueshifted H$\alpha$ absorption with at most a weak emission
  component (Bastian \& Mundt 1985). In addition, the far-red Ca~II
  triplet lines can either be in emission or absorption (Welty 1991;
  Welty et al. 1992).  During outburst, however, V1647~Ori showed
  strong H$\alpha$ emission with associated blueshifted absorption and
  the far-red \ion{Ca}{2} triplet lines are in emission.

\item The deep NIR CO overtone bandhead absorption features, typical
  of the four classical FUors (Reipurth \& Aspin 1997; Connelley \&
  Greene 2010), have not been observed in V1647~Ori, rather CO
  bandhead emission was seen soon after the first outburst.  As the
  outburst subsided, the CO bandhead region became relatively
  featureless other than the weak CO absorption occurring in the
  quiescent period between the first and second outburst phases (see
  Figure~\ref{nirspec2}).

\item The curving nebulous structure prototypical of classical FUors
  (Goodrich 1987) is replaced in V1647~Ori by an extensive monopolar
  structure, McNeil's Nebula (see Figure~\ref{optims}).

\end{itemize}

One possibility to explain the dichotomy is that the viewing geometry
of the young star/disk system plays a role in what spectral features
are observed.  High inclinations could easily result in selective
illumination and/or obscuration.  The telltale sign of a FUor, the
presence of curving nebulosity extending away from the star (Goodrich
1987 presented several good example images) could be explained in
terms of selective illumination of material located in either the
walls of an evacuated cavity created by an earlier large-scale
molecular outflow, or in the outer regions of the circumstellar disk.
An earlier molecular outflow is very likely the origin of the
monopolar nebula (McNeil's nebula) in V1647~Ori which, as we have
seen, has an estimated inclination (with respect to the plane of the
sky) of $\sim$29$\pm$14$^{\circ}$ (Acosta-Pulido et al.  2007).  The
opening angle of this cavity may also play a role in what nebulous
structures are illuminated and hence observed.  The lack of monopolar
or bipolar structure associated with the classical FUors may well
suggest that they are viewed at a much larger inclination (closer to
90$^{\circ}$).  This interpretation would be unacceptable if the four
classical FUors were the only FUors known, since statistically, this
would be highly improbable.  However, two sources postulated to be
FUors, Par~21 (Staude \& Neckel 1992; Kospal et al.  2008) and
HH381~IRS (Reipurth \& Aspin 1997; Magakian et al.  2011), both have
bright monopolar cavities seen in reflected light and have inferred
inclinations $<$30$^{\circ}$.  Other FUor-like objects\footnotemark
\footnotetext{FUor-like objects are those which possess many of the
  characteristics of classical FUors but for which the outburst was
  not documented. A current list of FUors and FUor-like objects is
  given in Reipurth \& Aspin (2010).}  also possess well-defined
outflow cavities e.g.  PP~13S (Sandell \& Aspin 1998; Aspin \& Sandell
2001), L1551~IRS5 (Mundt et al. 1985), and V2495~Cyg (also known as
the Braid Nebula Star, Movsessian et al.  2006).  This means that a
line-of-sight towards the four classical FUors with, say, {\it
  i}$>$70$^{\circ}$ is not as unlikely as one might expect.

\section{Conclusions}
Our conclusions are:

\begin{enumerate}

\item V1647~Ori has remained in an elevated photometric state since
  late 2008, which means that the current (second) eruption is
  now approaching 3~yrs old.

\item McNeil's Nebula has remained bright and has a morphology very
  similar to that seen during the 2004 and 2008 outbursts.

\item CO overtone bandhead emission, seen in 2004 close to the
  beginning of the first eruption, is no longer present, with the
  spectral region from 2.9--2.4~$\mu$m now being featureless.

\item Water vapor absorption bands seen in the 1--2.5~$\mu$m spectral
  region, which developed as the first eruption proceeded, remain
  strong.

\item H$\alpha$ still maintains a P~Cygni profile indicative of the
  presence of a strong star/disk wind.  The minimum in the absorption
  component is found to be at around --100~km~s$^{-1}$.  The H$\alpha$
  FW10\% value of 620~km~s$^{-1}$ implies that accretion rates are
  high although the value derived using the analysis of Natta et al.
  (2004), is at least a factor of 100 less than that obtained from the
  dereddened Br$\gamma$ emission line fluxes.

\item The accretion luminosity and rate in early 2011, derived from
  the dereddened Br$\gamma$ flux, are 14$\pm$4~L$_{\odot}$ and
  3.8$\pm$2$\times$10$^{-6}$~M$_{\odot}$~yr$^{-1}$, respectively.
  These are statistically similar to those found throughout the
  sources elevated photometric state.  Although much of the 2007
  fading of V1647~Ori was caused by the changes in line-of-sight
    extinction, we believe that the accretion luminosity and rate at
  that time declined by a factor of 2--3.

\end{enumerate}

We wait with anticipation to track the changes occurring in V1647~Ori
over the next few years.

\vspace{0.3cm}

{\bf Acknowledgments:} We are extremely grateful to the anonymous
referee for many constructive comments and suggestions on the
manuscript.  We wish to thank John Rayner for the IRTF engineering time
during which some of the above data were obtained.  We also thanks
Eric Volquardsen for taking those observations on our behalf.  We
thank the University of Hawai'i observers who obtained optical imaging
of V1647~Ori during their observing nights.  Based on observations
obtained at the Gemini Observatory, which is operated by the
Association of Universities for Research in Astronomy, Inc., under a
cooperative agreement with the NSF on behalf of the Gemini
partnership: the National Science Foundation (United States), the
Science and Technology Facilities Council (United Kingdom), the
National Research Council (Canada), CONICYT (Chile), the Australian
Research Council (Australia), Ministério da Ciência e Tecnologia
(Brazil) and SECYT (Argentina).  This publication makes use of data
products from the Two Micron All Sky Survey, which is a joint project
of the University of Massachusetts and the Infrared Processing and
Analysis Center/California Institute of Technology, funded by the
National Aeronautics and Space Administration and the National Science
Foundation.  {\it The authors wish to recognize and acknowledge the
  very significant cultural role and reverence that the summit of
  Mauna Kea has always had within the indigenous Hawaiian community.
  We are most fortunate to have the opportunity to conduct
  observations from this sacred mountain.}


\clearpage

\begin{deluxetable}{lclr}
\tablecaption{Observation Log\label{obslog}}
\tablewidth{0pt}
\tablehead{
\colhead{UT Date} & 
\colhead{JD} &
\colhead{Telescope/} & 
\colhead{Details} \\
\colhead{} &
\colhead{} &
\colhead{Instrument} &
\colhead{}}
\startdata
2008 Aug 31  & 2454709 & UH 2.2m/Tek       & optical imaging \\
2008 Oct 09  & 2454748 & UH 2.2m/Tek       & optical imaging \\
2009 Jan 23  & 2454853 & UH 2.2m/Tek       & optical imaging \\
2009 Sep 11  & 2455084 & UH 2.2m/Tek       & optical imaging \\
2010 Jan 03  & 2455198 & Gemini-North/GMOS & optical imaging \\
2011 Feb 02  & 2455593 & Gemini-North/GMOS & optical imaging  \\
2011 Feb 02  & 2455593 & Gemini-North/GMOS & optical spectroscopy \\
2011 Feb 06  & 2455597 & Gemini-North/NIRI & NIR imaging \\
2011 Feb 15  & 2455606 & Gemini-North/NIFS & NIR IFU spectroscopy \\
2011 Apr 19  & 2455671 & IRTF/SpeX         & NIR spectroscopy \\
\enddata

\end{deluxetable}

\clearpage

\begin{deluxetable}{lr}
\tablecaption{Optical Photometry of V1647~Ori\label{optphot}}
\tablewidth{0pt}
\tablehead{
\colhead{UT Date} & 
\colhead{r'} \\
\colhead{} &
\colhead{(mags)}}
\startdata
2004 Mar 04 & 17.91$\pm$0.09 \\
2007 Feb 22 & 23.26$\pm$0.15 \\
2008 Aug 31 & 16.91$\pm$0.15 \\
2008 Sep 22 & 17.53$\pm$0.07 \\
2008 Oct 09 & 17.53$\pm$0.08 \\
2009 Jan 23 & 17.38$\pm$0.12 \\
2009 Sep 11 & 17.31$\pm$0.12 \\
2010 Jan 03 & 17.51$\pm$0.06 \\
2011 Feb 02 & 17.77$\pm$0.04 \\
\enddata

\end{deluxetable}

\clearpage

\begin{deluxetable}{lrrrrr}
\rotate
\tablecaption{NIR Photometry of V1647~Ori\label{nirphot}}
\tablewidth{0pt}
\tablehead{
\colhead{Filter/Color} & 
\colhead{UT 2004 Mar 4/9\tablenotemark{a}} &
\colhead{2007 Feb 22\tablenotemark{b}} &
\colhead{2008 Sep 22\tablenotemark{c}} &
\colhead{2011 Feb  2\tablenotemark{d}} &
\colhead{2011 Apr 19\tablenotemark{e}} \\
\colhead{} &
\colhead{(mags)} &
\colhead{(mags)} &
\colhead{(mags)} &
\colhead{(mags)} &
\colhead{(mags)}}
\startdata
r'     & 17.91$\pm$0.09 & 23.26$\pm$0.15 & 17.53$\pm$0.07 & 17.77$\pm$0.04 & -- \\
J      & 11.13$\pm$0.10 & 14.72$\pm$0.10 & 10.86$\pm$0.07 & 11.22$\pm$0.06 & 10.85$\pm$0.10 \\
H      &  9.09$\pm$0.07 & 11.90$\pm$0.10 &  9.00$\pm$0.07 &  9.22$\pm$0.02 &  8.87$\pm$0.10 \\
K'     &  7.48$\pm$0.09 & 10.14$\pm$0.10 &  7.74$\pm$0.07 &  7.81$\pm$0.03 &  7.58$\pm$0.10 \\
L'     &  --            &  7.62$\pm$0.10 &  5.76$\pm$0.10 &  5.58$\pm$0.10 & -- \\
J--H   &  2.04$\pm$0.12 &  2.82$\pm$0.14 &  1.86$\pm$0.10 &  2.00$\pm$0.07 &  1.98$\pm$0.14 \\
H--K'  &  1.61$\pm$0.11 &  1.76$\pm$0.14 &  1.26$\pm$0.10 &  1.41$\pm$0.04 &  1.29$\pm$0.14 \\
K'--L' &  --            &  2.52$\pm$0.14 &  1.98$\pm$0.14 &  2.23$\pm$0.14 & -- \\
\enddata

\tablenotetext{a}{r' data from Aspin \& Reipurth (2009).  J, H, K'
  data from Acosta-Pulido et al. (2007).}

\tablenotetext{b}{Data presented in Aspin, Beck, \& Reipurth (2008).}

\tablenotetext{c}{Data presented in Aspin et al. (2009).}

\tablenotetext{d}{Data from this paper (Gemini/NIRI.)} 

\tablenotetext{e}{Data from this paper (IRTF/SpeX, SXD only).} 

\end{deluxetable}

\clearpage

\begin{deluxetable}{lccccc}
\tablecaption{Optical Spectral Features\label{optfluxes}}
\tablewidth{0pt}
\tablehead{
\colhead{Line} & 
\colhead{$\lambda$} & 
\colhead{EW\tablenotemark{a}} &
\colhead{FWHM\tablenotemark{b}} &
\colhead{Line Flux\tablenotemark{c}} &
\colhead{Continuum\tablenotemark{c}} \\
\colhead{} & 
\colhead{(\AA)} & 
\colhead{(\AA)} &
\colhead{(km~s$^{-1}$)} &
\colhead{(ergs~cm$^{-2}$~s$^{-1}$)} &
\colhead{(ergs~cm$^{-2}$~s$^{-1}$~\AA$^{-1}$)}}
\startdata
\multicolumn{6}{c}{UT 2011 Feb 2} \\
$[$\ion{O}{1}$]$ & 6300 & --0.9  & 170 & 1.19(--16) & 1.32(--16) \\
H$\alpha$        & 6563 & --32.6 & 310\tablenotemark{d} & 6.72(--15) & 2.02(--16) \\
\ion{O}{1}       & 7773 & +2.3   & 270 & 1.60(--15) & 7.04(--16)  \\
\ion{Fe}{2}      & 8228 & +0.8   & 190 & 4.42(--16) & 7.37(--16) \\
\ion{Fe}{1}      & 8388 & --0.6  & 170 & 4.75(--16) & 7.91(--16) \\
\ion{O}{1}       & 8446 & --0.5  & 240 & 5.20(--16) & 8.17(--16) \\
\ion{Ca}{2}      & 8498 & --8.6  & 160 & 7.09(--15) & 8.15(--16) \\
\ion{Fe}{1}      & 8516 & --0.5  & 190 & 5.31(--16) & 8.25(--16) \\
\ion{Ca}{2}      & 8543 & --9.0  & 160 & 7.76(--15) & 8.41(--16) \\
\ion{Ca}{2}      & 8663 & --7.7  & 160 & 6.40(--15) & 8.47(--16) \\
\multicolumn{6}{c}{ } \\
\multicolumn{6}{c}{UT 2011 Apr 19\tablenotemark{e}} \\
\ion{Ca}{2}      & 8498 & --13.0  & 330 & 1.66(--14) & 1.24(--15) \\
\ion{Ca}{2}      & 8543 & --11.0  & 330 & 1.40(--14) & 1.31(--15) \\
\ion{Ca}{2}      & 8663 & --8.0   & 330 & 9.59(--15) & 1.43(--15) \\
\enddata
\tablenotetext{a}{Equivalent widths have associated uncertainties of
  $\pm$0.2~\AA.}  

\tablenotetext{b}{Full-Width Half Maximum line width. These values
  have associated uncertainties of $\pm$10~km~s$^{-1}$.}

\tablenotetext{c}{Number in parentheses is exponent e.g. 9.24(--17)
  means 9.24$\times$10$^{-17}$.}

\tablenotetext{d}{FWHM measured on the model fit to the observed
  H$\alpha$ profile.}

\tablenotetext{e}{Data from IRTF/SpeX spectrum that extends down to
  8000~\AA.}
\end{deluxetable}

\clearpage

\begin{deluxetable}{lrrrrr}
\tabletypesize{\scriptsize}
\tablecaption{H$\alpha$ Velocity Structure\label{ha}}
\tablewidth{0pt}
\tablehead{
\colhead{Feature} &
\colhead{UT 2004 Mar 10\tablenotemark{a}} &
\colhead{2007 Feb 21\tablenotemark{b}} &
\colhead{2008 Sep 22\tablenotemark{c}} &
\colhead{2011 Feb  2\tablenotemark{d}} &
\colhead{2010 Oct  7\tablenotemark{e}} \\
\colhead{} &
\colhead{V1647~Ori} &
\colhead{V1647~Ori} &
\colhead{V1647~Ori} &
\colhead{V1647~Ori} &
\colhead{FU~Ori}}
\startdata
\multicolumn{6}{c}{H$\alpha$ Emission} \\
V$_{max}$~(km~s$^{-1}$)\tablenotemark{f}    &     0\tablenotemark{g} &  +120 &     0 &  --35 & --60 \\
FW10\%~(km~s$^{-1}$)\tablenotemark{h}       &   530\tablenotemark{i} &   574 &   654 &   615 &  580 \\
FWHM$_{em}$~(km~s$^{-1}$)\tablenotemark{j}  &   240\tablenotemark{i} &   345 &   340 &   310 &  290 \\
 & & & & \\
\multicolumn{6}{c}{H$\alpha$ Absorption} \\
V$_{min}$~(km~s$^{-1}$)\tablenotemark{k}    & --240\tablenotemark{g} &  +70  & --385 & --110 & --60 \\
FWHM$_{abs}$~(km~s$^{-1}$)\tablenotemark{l} &   250\tablenotemark{i} &  215  &   485 &   205 &  177 \\
\enddata

\tablenotetext{a}{Data taken soon after the first outburst.  Line
  profile best-fit using a Voigt function.}

\tablenotetext{b}{Data taken between the first and second outburst.
  Line profile best-fit using a Gaussian function.}

\tablenotetext{c}{Data taken soon after the second outburst. Line
  profile best-fit using a Voigt function.}

\tablenotetext{d}{Data taken in 2011 February when the source was
  still optically bright. Line profile best-fit using a Voigt
  function.}

\tablenotetext{e}{Data taken in 2010 October.  Line profile best-fit
  using a Gaussian function.}

\tablenotetext{f}{Velocity of maximum H$\alpha$ emission (with respect
  to the nominal H$\alpha$ wavelength of 6562.8~\AA) measured from the
  best-fit model.}

\tablenotetext{g}{Typical uncertainties in quoted velocities are
  $\pm$30~km~s$^{-1}$.}

\tablenotetext{h}{Full-Width 10\% intensity of emission line measured
  on the model profile.}

\tablenotetext{i}{FW10\%, and FWHM velocities have been deconvolved
  using a Gaussian with FWHM of an unresolved arc line
  (i.e., 130~km~s$^{-1}$).}

\tablenotetext{j}{Full-Width Half Maximum of best-fit emission
  Gaussian measured on the model profile.}

\tablenotetext{k}{Velocity of the minimum in the blueshifted
  absorption component (with respect to the nominal H$\alpha$
  wavelength of 6562.8~\AA) determined from the (observed--model)
  profile.}

\tablenotetext{l}{Full-Width Half Maximum of the (observed--model)
  absorption profile.}

\end{deluxetable}

\clearpage

\begin{deluxetable}{lcccc}
\tablecaption{NIR Spectral Features\label{nirfluxes}}
\tablewidth{0pt}
\tablehead{
\colhead{Line} & 
\colhead{$\lambda$} & 
\colhead{EW\tablenotemark{a}} &
\colhead{Line Flux\tablenotemark{b}} &
\colhead{Continuum\tablenotemark{b}} \\
\colhead{} & 
\colhead{($\mu$m)} & 
\colhead{(\AA)} &
\colhead{(ergs~cm$^{-2}$~s$^{-1}$)} &
\colhead{(ergs~cm$^{-2}$~s$^{-1}$~\AA$^{-1}$)}}
\startdata
\multicolumn{5}{c}{UT 2011 Feb 15} \\
Pa$\beta$       & 1.282  & --7.8  & 8.83(--14) & 1.14(--14) \\
Br$\gamma$      & 2.167  & --5.5  & 2.02(--13) & 3.64(--14) \\
\multicolumn{5}{c}{ } \\
\multicolumn{5}{c}{UT 2011 Apr 19} \\
\ion{He}{1}     & 1.082  &  +5.6  & 3.45(--14) & 6.75(--15) \\
Pa$\gamma$      & 1.094  & --1.0  & 7.78(--15) & 7.21(--15) \\
                & 1.257  & --1.0  & 9.72(--15) & 1.43(--14) \\
Pa$\beta$       & 1.282  & --6.0  & 1.01(--13) & 1.62(--14) \\
Br$\gamma$      & 2.167  & --4.0  & 1.54(--13) & 4.52(--14) \\
\enddata

\tablenotetext{a}{Equivalent widths have associated uncertainties of 
$\pm$0.2~\AA.}

\tablenotetext{b}{Number in parentheses is exponent e.g. 3.45(--14) 
means 3.45$\times$10$^{-14}$.}

\end{deluxetable}

\begin{deluxetable}{lcccccc}
\tablecaption{Accretion Luminosity and Rate Estimates from NIR Line Fluxes\label{pabbrg}}
\tablewidth{0pt}
\tablehead{
\colhead{UT} & 
\colhead{A$_V$\tablenotemark{a}} &
\colhead{Pa$\beta$/Br$\gamma$\tablenotemark{b}} &
\colhead{L$_{acc}$\tablenotemark{c}} &
\colhead{\.M$_{acc}$\tablenotemark{c}} &
\colhead{L$_{acc}$\tablenotemark{d}} &
\colhead{\.M$_{acc}$\tablenotemark{d}} \\
\colhead{Date} &
\colhead{(mags)} &
\colhead{} &
\colhead{(L$_{\odot}$)} &
\colhead{(10$^{-6}$~M$_{\odot}$~yr$^{-1}$)} &
\colhead{(L$_{\odot}$)} &
\colhead{(10$^{-6}$~M$_{\odot}$~yr$^{-1}$)}}
\startdata
 & & & \multicolumn{2}{c}{Pa$\beta$} & \multicolumn{2}{c}{Br$\gamma$} \\
2004 Mar  9\tablenotemark{e}  &  8$\pm$2 & 2.2$\pm$0.3 & 5.1$\pm$2 & 1.3$\pm$0.6 (--5.88\tablenotemark{f}) & 20$\pm$6    & 5.3$\pm$3 (--5.28) \\
2007 Feb 22\tablenotemark{g}  & 19$\pm$2 & 4.9$\pm$0.3 & 3.4$\pm$2 & 0.9$\pm$0.3 (--6.06) &  5$\pm$2    & 1.2$\pm$1 (--5.92) \\
2008 Aug 31\tablenotemark{h}  &  8$\pm$2 & 1.4$\pm$0.3 & 2.4$\pm$1 & 0.6$\pm$0.2 (--6.19) & 16$\pm$5    & 4.0$\pm$3 (--5.39) \\
2011 Feb 15\tablenotemark{i}  &  8$\pm$2 & 1.3$\pm$0.3 & 2.4$\pm$1 & 0.6$\pm$0.2 (--6.20) & 17$\pm$5    & 4.5$\pm$2 (--5.35) \\
2011 Apr 19\tablenotemark{i}  &  8$\pm$2 & 1.9$\pm$0.3 & 2.8$\pm$1 & 0.7$\pm$0.2 (--6.13) & 12$\pm$4    & 3.2$\pm$2 (--5.50) \\
\enddata

\tablenotetext{a}{Visual extinction used to deredden line fluxes.
  Derived from dereddening J-H vs. H-K' colors.}

\tablenotetext{b}{Ratio of dereddened Pa$\beta$ and Br$\gamma$ line
  fluxes. Case~B recombination theory gives a value of $\sim$6.}

\tablenotetext{c}{Determined from the dereddened Pa$\beta$ line flux.}

\tablenotetext{d}{Determined from the dereddened Br$\gamma$ line flux.}

\tablenotetext{e}{Fluxes from Vacca et al. (2004).}

\tablenotetext{g}{Fluxes from Aspin, Beck, \& Reipurth (2008).}

\tablenotetext{f}{Values in parentheses are log$_{10}$ of the preceding accretion rates.}

\tablenotetext{h}{Fluxes from Aspin et al. (2009).}

\tablenotetext{i}{Fluxes from this paper.}

\end{deluxetable}

\clearpage
\begin{deluxetable}{lcr}
\tablecaption{Accretion Luminosity and Rate Estimates from H$\alpha$ FW10\% values\label{natta}}
\tablewidth{0pt}
\tablehead{
\colhead{UT} & 
\colhead{FW10\%\tablenotemark{a}} &
\colhead{\.M$_{acc}$\tablenotemark{b}} \\
\colhead{Date} &
\colhead{(km~s$^{-1}$)} &
\colhead{(10$^{-8}$~M$_{\odot}$~yr$^{-1}$)}}
\startdata
\multicolumn{3}{c}{V1647~Ori} \\
2004 Mar 10\tablenotemark{c} & 530 &  2(0.4--8)\tablenotemark{d} \\
2007 Feb 21\tablenotemark{e} & 574 &  5(1--24) \\
2008 Sep 22\tablenotemark{f} & 654 & 30(5--163) \\
2011 Feb  2\tablenotemark{g} & 615 & 12(2--64) \\
 & & \\
\multicolumn{3}{c}{FU~Ori} \\
2011 Oct  7\tablenotemark{g} & 580 & 5(1--28) \\
\enddata

\tablenotetext{a}{H$\alpha$ Full-Width 10\% intensity measured on the
  best-fit model profile.}

\tablenotetext{b}{Accretion rate derived using analysis of Natta et
  al. (2004).}

\tablenotetext{c}{From Aspin \& Reipurth (2009).}

\tablenotetext{d}{Values in parentheses are the range of value defined
  by the uncertainties in the Natta et al. (2004) relationship.}

\tablenotetext{e}{Data from Aspin, Beck, \& Reipurth (2008).}

\tablenotetext{f}{Data from Aspin et al. (2009).}

\tablenotetext{g}{Data from this paper.}

\end{deluxetable}

\clearpage

\clearpage

\clearpage

\clearpage

\clearpage

\clearpage

\clearpage

\clearpage

\end{document}